{
\documentclass{article}
 \relax
\setlength{\oddsidemargin}{.25in}
\setlength{\textwidth}{5.25in}
\setlength{\topmargin}{.25in}
\setlength{\textheight}{7.5in}
\usepackage{graphics}

\begin{document}

\title{Influence On The Physical Universe By Wormhole Generated Extra Dimensional Space
}         
\author{A. L. Choudhury   \\
Department of Chemistry and Physics  \\ Elizabeth City State University  \\
Elizabeth City, NC 27909 \\ email:alchoudhury@mail.ecsu.edu }      
\date{August 19, 2003}          
\maketitle

\begin{abstract}
   {  We use the Gu-Huang model for a special case when the universe is seven dimensional. In the core of the extra dimension we place a modified Gidding-Strominger wormhole. This wormhole is separated by a thin wall from the extra dimensional space. The wormhole content is assumed to satisfy the adiabatic gas law. The wormhole pressure  penetrates into the extra dimension. We then solve the Einstein equation assuming that the real universe and the extra dimension contains only inextendable fluid with negligible local pressure. We show the physical universe expands because the time dependent Hubble parameter is positive. Under certain condition the deacceleration parameter, $q_0$ is also positive. But the most significant outcome of our investigation is the fact that $q_0$ fluctuates. If we can detect by observation that the acceleration fluctuates, our model will be an alternative explanation of the expanding universe without the help of dark energy. We could however link dark energy with work done by wormhole pressure.       }
\end{abstract}

\section{Introduction}
   {$\;$$\;$Recently Gu and Huang [1] introduced an interesting model with extra dimensions. They showed that their model incorporates an accelerating universe under certain special conditions. They considered (3+n+1)-dimensional space time in which both the ordinary 3-space and the extra space are homogeneous and isotropic. They restrained the radii of the extra dimension between 0 and 1.
$\;$$\;$Here we simplify their model by restraining it to only  three extra dimensions. However, we place a modified Gidding-Strominger [2,3] wormhole in the middle of the extradimensional universe. It has been shown by Choudhury and Pendharkar [4] and also by Choudhury [5] that if we assume the content of the wormhole expands satisfying adiabatic gas law, its pressure becomes time dependent. Separating the wormhole and extradimensional space by a thin but flexible wall which allows the pressure to be conveyed into extra dimensional space. The time dependent pressure would then influence the expansion of the real physical universe we live in.
$\;$$\;$The time dependent pressure generates the Hubble expansion and under certain restrictions allows the physical universe to accelerate. One of the striking features of our model is the fact that the deacceleration parameter fluctuates with time staying always negative. This variation of the deacceleration parameter can be tested by sensitive observation.} 
{$\;$$\;$We assume that the unobservable extra dimension has a definite expansion mode. If future observation shows that such fluctuation is real, this assumption will have a strong foothold.
$\;$$\;$In section 2, we repeat some properties of the wormhole core. We show how the pressure in the wormhole is generated in section 3. In section 4, by following Gu and Huang we introduce a seven dimensional space which is homogeneous and isotropic. We also mention there that the real world and the extra dimensions contain fluid under pressure. From Einstein's equation we derive the equations satisfied by the scale factors. We also obtain there the Hubble parameter and show that it is always positive.
$\;$$\;$In section 5 we derive the deacceleration parameter $q_0$. We also show that under certain restriction, $q_0$ can always be made less than zero. One of the most important outcome of our result is the fact that the $q_0$ fluctuates. 
$\;$$\;$In section 6 we discuss our result.   }
\section{The unphysical wormhole core}
{$\;$$\;$The wormhole core which generates pressure on the Gu and Huang [1] universe with with extra dimension is assumed to be obtained from the modified Gidding-Srominger model[2,3] as summerized in the papers Choudhury and Pendharkar [4], and Choudhury [5]. For the unphysical inner core we start to construct the model in an Euclidean space. After getting the solution of the wormhole scale factor we switch over to the Lorentz space by substituting 't' in Euclidean space solutions with 'it'. We assume that such substitution is mathematically permissible. The core action in the Euclidean space is given as follows:    
\begin {equation}
S_E=\int {{d^4}x {{L_G}^c}(x)}+\int {{d^4}x {{L_{SA}}^c}(x)}=S_G+S_A,
\end{equation}
{where}
\begin {equation}
{{L_G}^c}(x)=\frac{{\surd g^c}{R(g^c)}} {2{\kappa}^2},
\end{equation}
{and}
\begin{equation}
{{L_{SA}}^c} (x)={\surd {g^c}}[{\frac1 2}(\bigtriangledown \Phi)^2+{g_p}^2 {\Phi}^2 Exp(\beta {\Phi}^2)]{H_{\mu\nu\rho}}^2.
\end{equation}
{In Eqs.(1) through (3)the suffix c stands for the core. The axion field is given by the relation }
\begin{equation}
H_{\mu\nu\rho}=\frac n {{g_p}^2 {a^3 (t)}} \varepsilon_{\mu\nu\rho},
\end{equation}
{ from which we get}
\begin{equation}
H^2=\frac{6D} {{g_p}^2 {a^6}},
\end{equation}
{with}
\begin{equation}
D=\frac{n^2} {{g_p}^2}
\end{equation}
{The space-time interval in Euclidean space is given by}
\begin {equation}
ds^2=dt^2+a^2(t)(d\chi^2+sin^2 \chi d\theta^2+sin^2 \chi sin^2 \phi d\phi^2).
\end{equation}
{The variation of $g_{\mu\nu}^c$ in the core leads to the following equation}
\begin{equation}
R_{\mu\nu}^c-{\frac1 2}g_{\mu\nu}^c R^c=\kappa^2[\nabla_\mu \Phi \nabla_\nu \Phi-({\frac1 2}(\nabla\Phi)^2) g_{\mu\nu}^c+{g_p}^2 Exp(\beta {\Phi}^2)({H_\mu}^\alpha\gamma H_\nu\alpha\gamma- {\frac1 6} g_{\mu\nu}^c H_{\alpha\beta\gamma} ^2)]
\end{equation}
{For the $\Phi$-variation inside the core the equation of motion of $\Phi$ yields}
\begin{equation}
{\nabla}^2 \Phi-2{{g_p}^2\beta}\Phi{Exp(\beta\Phi^2)}{H_{\alpha\beta\rho}}^2=0.
\end{equation}
{ The Hamiltonian constraint yields}
\begin{equation}
(\frac1 a \frac{da} {dt})^2-\frac1 {a^2}=\frac{\kappa^2} 3 [{\frac1 2}(\frac{d\Phi} {dt})^2-6Exp(\beta\Phi^2) \frac{n^2} {g_p^2 a^6}].
\end{equation}
{The dynamical equation yields }
\begin{equation}
{\frac d {dt}} ({\frac1 a }{\frac{da} {dt}})+\frac1 {a^2}=-\kappa^2[(\frac{d\Phi} {dt})^2-12 Exp(\beta\Phi^2) {\frac{n^2} {g_p^2 a^6}}].
\end{equation}
{ Now we substitute  a new variable $\tau$ defined by the relation}
\begin{equation}
d\tau=a^{-3} dt,
\end{equation}
{Combining these equations we can derive the equation satisfied by the scale function for the wormhole core as }

\begin{equation}
{(\frac1 a }{\frac{da} {dt}})^2-a^4+{a_c}^4=0,
\end{equation}
{where}
\begin{equation}
a_c=\surd(\frac{\kappa^2C_o} 2)
\end{equation}
{and $C_o$ is a constant. }
{$\;$$\;$The solution of the Eq.(13) has been obtained by Gidding and Strominger and is given by}
\begin{equation}
a^2(\tau)={a_c}^2\surd(sec(2{a_c}^2\tau)).
\end{equation}
{The scale factor can now be of two possible forms }
\begin{equation}
a(\tau)=\pm a_c (sec(2{a_c}^2\tau))^{\frac1 4},
\end{equation}
{specified by the signs. Both the solutions stand on equal footing.}
\section{Pressure generated by the wormhole}
{$\;$$\;$ Following suggestions originated and further developed by Choudhury and Pendharkar [4] we assume that the wormhole is in a gaseous state satisfying the adiabatic gas law }   
\begin{equation}
P_w V_w ^\gamma=Constant=B_1,
\end{equation}
{where $\gamma$  is a constant. The volume of the wormhole can be shown to be proportional to $a^3 (\tau)$ . The pressure can be expressed as }
\begin{equation}
P_w={(\pm 1)^{-3\gamma}}B{a_c}^{-3\gamma}[sec(2{a_c}^2 \tau)]^{-(3\gamma/4)},
\end{equation}
{where B is a new constant. Switching over to real time from by going $t\rightarrow it$ we get from Eqs. (12) and (15) }
\begin{equation}
d\tau={\pm}i{a_c}^{-3}{\frac{dt}  {[cosh{\frac{2t} {a_c}}]^{3/4}}}.
\end{equation}
{For large t we neglect the negative exponential in cosh term and obtain for $\tau$ the value}
\begin{equation}
\tau={\pm}i{\frac{2^{\frac1 4}} 3} {a_c}^{-2}Exp({\frac{3t} {2{a_c}}}).
\end{equation}

{Substituting this value of $\tau$ Eq.(18) a($\tau$) changes into}
\begin{equation}
a(\tau)={\pm}{a_c}[cosh(f(t) )]^{-\frac1  4},
\end{equation} 
{where we have set $f(t)={\frac{2^{\frac5 4}} 3}Exp[{\frac{3t} {2a_c}}]$.}
{Following Choudhury and Pendharkar [4] we can convert Eq.(18) into}

\begin{equation}
P_w={\pm }B{a_c}^{-3\gamma}[cosh(f(t) )]^{(3\gamma/4)}.
\end{equation}

\section{ Seven dimensional universe and Hubble parameter}
{$\;$$\;$We here introduce a seven dimensional physical universe, a special case of the Gu and Huang model[1]. This space we assume to be homogeneos and isotropic Robertson-Walker space. The interval is given by the relation}  

\begin{equation}
d {s_P}^2 = -d t^2 + d {s_\alpha}^2+ d {s_\beta}^2
\end{equation}
{where }
\begin{equation}
d {s_\alpha}^2={{\alpha^2}(t)}[\frac{d{r_\alpha}^2} {1-{k_\alpha}{r_\alpha}^2} + {{r_\alpha}^2}d {\theta_\alpha}^2 + {{r_\alpha}^2}{sin^2}{\theta_\alpha} d{\phi_\alpha}^2],
\end{equation}
{and}
\begin{equation}
d {s_\beta}^2={{\beta^2}(t)}[\frac{d{r_\beta}^2} {1-{k_\beta}{r_\beta}^2} + {{r_\beta}^2}d {\theta_\beta}^2 + {{r_\beta}^2}{sin^2}{\theta_\beta} d{\phi_\beta}^2]
\end{equation}

{We have chosen c=1. The Einstein equation is as follows}

\begin{equation}
G_{\mu\nu}=R_{\mu\nu}-{\frac1 2}g_{\mu\nu} R  = - {8 \pi G}{T_{\mu\nu}}.
\end{equation}

{In the above equation $\mu$ and $\nu$ run from 0 through 6. The tensor $T_{\mu\nu}$ can be defined as}
\begin{equation}
T_{\mu\nu}= G {T_{\mu\nu}}^{(a)}+ G' {T_{\mu\nu}}^{(b)},
\end{equation}
{where}
\begin{equation}
{T_{\mu\nu}}^{(i)}= p_i {g_{\mu\nu}}^{(i)}+(p_i +\rho_i) {U_\mu}^i {U_\nu}^i,
\end{equation}
{with i=a and b.
In the above expressions}
\begin{equation}
{g_{\mu\nu}}^{(a)}=-1,  for  \mu=\nu=0;
\end{equation}
\begin{equation}
{g_{\mu\nu}}^{(a)}=1,  for  \mu=\nu=1,2,3;
\end{equation}
\begin{equation}
{g_{\mu\nu}}^{(a)}=0,  for  \mu\not=\nu=1,2,3;
\end{equation}
{and}
\begin{equation}
{g_{\mu\nu}}^{(a)}=0,  for  \mu=\nu=4,5,6.
\end{equation}
{$\;$$\;$Similarly}
\begin{equation}
{g_{\mu\nu}}^{(b)}=-1,  for  \mu=\nu=0;
\end{equation}
\begin{equation}
{g_{\mu\nu}}^{(b)}=-1,  for  \mu=\nu=4,5,6;
\end{equation}
\begin{equation}
{g_{\mu\nu}}^{(b)}=0,  for  \mu\not=\nu=1,2,3;
\end{equation}
{and}
\begin{equation}
{g_{\mu\nu}}^{(a)}=0,  for  \mu=\nu=1,2,3.
\end{equation}
{$\;$$\;$The equation}
\begin{equation}
G_{tt}=-8\pi(G{T_{tt}}^(a)+G'{T_{tt}}^(b)
\end{equation}
{turns into the form}
\begin{equation}
3[\frac{(\dot\alpha(t))^2+k_\alpha } {(\alpha(t))^2}]+\frac{(\dot\beta(t))^2+k_\beta } {(\beta(t))^2}+2{\frac{\dot\alpha(t)} {\alpha(t)}} {\frac{\dot\beta(t)} {\beta(t)}}=-8\pi(G{\rho_a}+G'{\rho_b})=-8\pi\rho.
\end{equation} 
{ Similarly we get for the ${r_a}$-${r_a}$ component}
\begin{equation}
2\frac{\ddot{\alpha}(t)}  {\alpha(t)}+3\frac{\ddot{\beta}(t)} {\beta(t)}+3\frac{(\dot\alpha(t))^2+k_\alpha } {(\alpha(t))^2}+3\frac{(\dot\beta(t))^2+k_\beta } {(\beta(t))^2}+6{\frac{\dot\alpha(t)} {\alpha(t)}} {\frac{\dot\beta(t)} {\beta(t)}}= 8{\pi} G{p_a}.
\end{equation}
{Both $\theta_a$-$\theta_a$ and $\phi_a$-$\phi_a$ component equations are given by}
\begin{equation}
2\frac{\ddot{\alpha}(t)}  {\alpha(t)}+3\frac{\ddot{\beta}(t)} {\beta(t)}+\frac{(\dot\alpha(t))^2+k_\alpha } {(\alpha(t))^2}+3\frac{(\dot\beta(t))^2+k_\beta } {(\beta(t))^2}+3{\frac{\dot\alpha(t)} {\alpha(t)}} {\frac{\dot\beta(t)} {\beta(t)}}= 8{\pi}G{p_a}.
\end{equation}
{$\;$$\;$For ${r_b}$-${r_b}$ component we get}
\begin{equation}
3\frac{\ddot{\alpha}(t)}  {\alpha(t)}+2\frac{\ddot{\beta}(t)} {\beta(t)}+3\frac{(\dot\alpha(t))^2+k_\alpha } {(\alpha(t))^2}+3\frac{(\dot\beta(t))^2+k_\beta } {(\beta(t))^2}+6{\frac{\dot\alpha(t)} {\alpha(t)}} {\frac{\dot\beta(t)} {\beta(t)}}= 8{\pi} G'{p_b}.
\end{equation}
{For $\theta_b$-$\theta_b$ and $\phi_b$-$\phi_b$ components we obtain}
\begin{equation}
3\frac{\ddot{\alpha}(t)}  {\alpha(t)}+2\frac{\ddot{\beta}(t)} {\beta(t)}+3\frac{(\dot\alpha(t))^2+k_\alpha } {(\alpha(t))^2}+\frac{(\dot\beta(t))^2+k_\beta } {(\beta(t))^2}+3{\frac{\dot\alpha(t)} {\alpha(t)}} {\frac{\dot\beta(t)} {\beta(t)}}= 8{\pi} G'{p_b}.
\end{equation}
{ In this paper we set $k_a$=$k_b$=0. The above equations then changes into the form}
\begin{equation}
\frac{(\dot\alpha(t))^2 } {(\alpha(t))^2}+\frac{(\dot\beta(t))^2 } {(\beta(t))^2}+2{\frac{\dot\alpha(t)} {\alpha(t)}} {\frac{\dot\beta(t)} {\beta(t)}}=-{\frac8 3}\pi\rho,
\end{equation} 
\begin{equation}
2\frac{\ddot{\alpha}(t)}  {\alpha(t)}+3\frac{\ddot{\beta}(t)} {\beta(t)}+3\frac{(\dot\alpha(t))^2 } {(\alpha(t))^2}+3\frac{(\dot\beta(t))^2 } {(\beta(t))^2}+6{\frac{\dot\alpha(t)} {\alpha(t)}} {\frac{\dot\beta(t)} {\beta(t)}}= 8{\pi} G{p_a},
\end{equation}
\begin{equation}
3\frac{\ddot{\alpha}(t)}  {\alpha(t)}+2\frac{\ddot{\beta}(t)} {\beta(t)}+3\frac{(\dot\alpha(t))^2 } {(\alpha(t))^2}+\frac{(\dot\beta(t))^2 } {(\beta(t))^2}+3{\frac{\dot\alpha(t)} {\alpha(t)}} {\frac{\dot\beta(t)} {\beta(t)}}= 8{\pi} G'{p_b},
\end{equation}
\begin{equation}
2\frac{\ddot{\alpha}(t)}  {\alpha(t)}+3\frac{\ddot{\beta}(t)} {\beta(t)}+\frac{(\dot\alpha(t))^2 } {(\alpha(t))^2}+3\frac{(\dot\beta(t))^2 } {(\beta(t))^2}+3{\frac{\dot\alpha(t)} {\alpha(t)}} {\frac{\dot\beta(t)} {\beta(t)}}= 8{\pi}G{p_a},
\end{equation}
{and}
\begin{equation}
3\frac{\ddot{\alpha}(t)}  {\alpha(t)}+2\frac{\ddot{\beta}(t)} {\beta(t)}+3\frac{(\dot\alpha(t))^2} {(\alpha(t))^2}+\frac{(\dot\beta(t))^2} {(\beta(t))^2}+3{\frac{\dot\alpha(t)} {\alpha(t)}} {\frac{\dot\beta(t)} {\beta(t)}}= 8{\pi} G'{p_b}.
\end{equation}
{$\;$$\;$Subtracting from Eq.(47) the Eq.(45) we get}
\begin{equation}
2\frac{(\dot\beta(t))^2 } {(\beta(t))^2}+3{\frac{\dot\alpha(t)} {\alpha(t)}}\frac{(\dot\beta(t))} {(\beta(t))}=0.
\end{equation} 
{Combining Eqs.(48) and (43) we get}
\begin{equation}
\frac{(\dot\alpha(t))^2 } {(\alpha(t))^2}+{\frac1 2}{\frac{\dot\alpha(t)} {\alpha(t)}}{\frac{\dot\beta(t)} {\beta(t)}}=-8{\pi}{\rho}.
\end{equation} 
{We have introduced extra dimensions to adjust the real universe with co-ordinates t, $r_a$,$\theta_a$,and $\phi_a$ . We assume that the space with extra dimensions expands with constant rate. We conjecture that the rate of expansion of the extra dimensions  satisfies the relation}
\begin{equation}
\frac{(\dot\beta(t)) } {(\beta(t))}= 4\Delta=H_b(t).
\end{equation}
{where $\Delta$ is a constant. The solution which $\beta(t)$ satisfies can be written as}
\begin{equation}
\beta(t)={\beta(0)}{e^{4\Delta t}}.
\end{equation}
{Eq.(49) then changes into the following form:}
\begin{equation}
\frac{(\dot\alpha(t))^2 } {(\alpha(t))^2}+2\Delta{\frac{\dot\alpha(t)} {\alpha(t)}}+8{\pi}{\rho}=0.
\end{equation}
{Writing } 
\begin{equation}
H_{\alpha}=\frac{\dot\alpha(t)} {\alpha(t)},
\end{equation}
{we get from Eq.(52)}
\begin{equation}
H_{\alpha}=-\Delta\pm\surd({\Delta}^2+8\pi\rho).
\end{equation}
{$\;$$\;$Since $H_\alpha$ is supposed to be the Hubble parameter of the real world with coordinates t,$r_a$,$\theta_a$,and $\phi_a$ to match the expanding universe, we discard the solution with the negative sign before the square root. Therefore we get for ${H_\alpha}(t)$}
\begin{equation}
H_{\alpha}=-\Delta+\surd({\Delta}^2+8\pi\rho).
\end{equation}
{Since G, G', $\rho_a$, $\rho_b$ are assumed to be positive we find that}
\begin{equation}
H_{\alpha}>0.
\end{equation}
{The time dependence of tthe Hubble parameter lies in the fact that the densities of the real and extra dimensions may depend on time.}
\section{The deaccelerating parameter of the real world}
{$\;$$\;$From the Eqs(44) and (45) we get}
\begin{equation}
\frac{\ddot{\alpha}(t)}  {\alpha(t)}-\frac{\ddot{\beta}(t)} {\beta(t)}=8\pi(G'p_b-Gp_a).
\end{equation}
{Due to the Eq.(50) we find}
\begin{equation}
\frac{\ddot{\alpha}(t)}  {\alpha(t)}= 16{\Delta}^2 + 8\pi(G'p_b-Gp_a).
\end{equation}
{Looking back to our construction of the universe, we assumed that the wormhole is placed at the center of the extra dimensions. The extra dimensional space is expanding at the rate of $H_b(t)$. The wormhole pressure is transferred through the wall from the wormhole to the extra dimension. If we now assume that there is no extra pressure except the pressure transferred through the wall to the extra dimensional space,then $p_b=p_w$. We get then the deacceleration parameter $q_0$ to be}
\begin{equation}
{q_0}(t_0)=-{\frac{\ddot{\alpha}(t)}  {\alpha(t)}}{\frac1 {{H_{\alpha}^2}(t_0)}}
=-[16{\Delta}^2 + 8\pi(G'p_b-Gp_a)]{\frac1 {{H_{\alpha}^2}(t_0)}},
\end{equation} 
{where $t_0$ represents present time (see Choudhury and Pendharkar[4]).

$\;$$\;$In Eq.(22) we make a special choice for the material of wormhole to satisfy the relation $\gamma=\frac4 3$. This assumption is highly speculative but has a strong predictive power.Under this assumption we get}
\begin{equation}
p_b  = B{{a_c}^{-4}}cosh^2(f(t) ).
\end{equation} 
{Therefore we get}
\begin{equation}
q_0(t_0)=-[16{\Delta}^2+8\pi[G'B{{a_c}^{-4}}cosh(f(t)) )-Gp_a]].
\end{equation}
{If the expansion rate $\Delta$ is assumed to satisfy a relation}
\begin{equation}
16{\Delta}^2-8\pi G p_a \geq 8\pi G' B{{a_c}^{-4}},
\end{equation}
{and we get}
\begin{equation}
q_0(t_0)\leq -8\pi G'B{{a_c}^{-8}}[1+cosh(f(t) )].
\end{equation}
{The above quantity $q_0$ is always negative. Therefore the universe is accelerating [6,7]. }  

\section{Concluding Remarks}
{$\;$$\;$Following Gu and Huang, we have constructed a model introducing extra dimensions. This model is a special case of their model where we have introduced only an extra three dimensions in addition to the physical space we live in. However we have incorporated an expanding modified Gidding-Strominger wormhole at the center of the extra dimension. This wormhole generates an adiabatic pressure . A flexible wall separates the wormhole from the extra dimensional space. This pressure influences the deacceleration parameter of our expanding universe. Introducing certain restriction on the parameters we have shown that the observational outcome of the accelerating universe can be reproduced. }

\section{References}
\begin{enumerate}
\item {Je-An Gu and W-Y. P. Huang, arXiv:astro-ph/0112565 v1 31 Dec 2001.}
\item { S. B. Giddings and A.  Strominger , Nucl. Phys. B 307, 854 (1988).}
\item { D. H. Coule and K. Maeda, Class. Quant. Grav. 7, 955 (1990).}
\item { L. Choudhury and H. Pendharkar, Hadronic J. 24, 275 (2001).}
\item { A. L. Choudhury, Hadronic J., 23, 581 (2000).}
\item { N. Bahcall, J. P. Ostriker, S. Perlmutter, and P. J. Steinhardt, Science, 284,1481 (1999).}
\item { C. Aremendariz Picon, V. Mukhanov and Paul J. Steinhardt: Essentials of k-Essence. ArXiv:astro-ph/0006373 (2000).}
\end{enumerate}    


\end{document}